\documentclass[10pt,leqno]{amsart}
\usepackage{graphicx}
\usepackage{indentfirst,csquotes}

\usepackage{amssymb,amsthm,amsmath}
\usepackage{xcolor,paralist,hyperref,titlesec,fancyhdr,etoolbox}

\titleformat{\section}
  {\normalfont\Large\bfseries}
  {\thesection}
  {1em}
  {}
\titlespacing*{\section}{0pt}{0ex}{0ex}

\titleformat{\subsection}
  {\normalfont\normalsize\bfseries}
  {\thesubsection}
  {1em}
  {}

\hypersetup{ colorlinks=true, linkcolor=black, filecolor=black, urlcolor=black }

\usepackage{lipsum}

\usepackage{minted}
\usepackage[numbers,sort&compress]{natbib}
\usepackage{float} 
\usepackage{hyperref} 
\usepackage{pgf-pie} 
\usepackage{marginnote} 
\usepackage{pgfplots} 
\usepgfplotslibrary{groupplots}
\pgfplotsset{compat=1.18}  
\usepackage{longtable} 

\usepackage{textcomp} 

\usepackage{appendix} 

\usepackage{subcaption} 

\setminted{breaklines=true}

\begin{document}
\title{Mapping Recent Shifts in Digital Art via Conference Discourse: AI, XR, the Metaverse, and Blockchain/NFTs (2021 - 2025)} 

\author[K]{Vasileios Komianos}
\date{\today}
\address{Department of Audio and Visual Arts, Ionian University,
  Corfu, Greece}
\email{vkomianos@ionio.gr}

\author[R]{Emmanuel Rovithis}
\date{\today}
\address{Department of Audio and Visual Arts, Ionian University,
  Corfu, Greece}
\email{emrovithis@ionio.gr}

\author[T]{Athanasios Tsipis}
\date{\today}
\address{Department of Digital Media and Communication, Ionian University, Argostoli, Greece}
\email{atsipis@ionio.gr}

\maketitle


\begin{abstract}
This paper presents an analysis of five years (2021–2025) of conference discourse across six digital art conferences, aiming to trace thematic shifts associated with the rapid development of emerging technologies, namely artificial intelligence (AI), immersive technologies (including XR and the metaverse), and blockchain technologies and non-fungible tokens (NFTs). The results indicate a marked increase in AI-related contributions, while immersive technologies maintain a relatively stable share of the discourse, and blockchain- and NFT-based works remain marginal. Overall, whereas immersive technologies and blockchain-related topics exhibit relative stability, AI shows a significant rise after 2022, emerging as a dominant theme within digital art conference discourse.
\end{abstract} 

\bigskip

\section{Introduction}
\noindent 
Digital art, as its name suggests, relies heavily on the use of digital technologies, which function as the primary creative tools of digital artists \citep{tykhoniuk2024analysis}. Based on the assumption that evolving digital technologies influence artistic creation, and that this influence is reflected in the literature produced both by artists themselves and by scholars examining digital art practices, this study aims to trace recent shifts in digital art through an analysis of conference discourse from the last five years, focusing primarily on artificial intelligence, and secondarily on immersive technologies (XR/VR/MR, metaverse) and applications, and blockchain technologies, including NFTs.

This study aims to outline the impact of evolving technologies by analyzing recent research efforts and artistic practices as presented in conference venues and their corresponding published proceedings. The decision to focus on conference proceedings was made based on a number of observations indicating that conferences may have not been thoroughly studied by the limited number of existing relevant systematic reviews.
The study aims to identify recent studies with similar objectives and then formulate its research questions and methodology based on the gaps and opportunities identified for further research. The proposed methodology includes the selection of conferences whose contributions are considered for analysis, the procedures applied to examine these contributions, and, finally, the presentation and discussion of the findings in relation to key research milestones and relevant technological and market developments. The results indicate that AI-related contributions are emerging as a major topic within the examined digital art conferences. Immersive technologies—including the family of extended reality (XR) technologies (VR/AR/MR) and the metaverse—constitute the second most prominent topic, following AI-related work, and have remained a relatively stable area of interest throughout the period under study. In contrast, contributions based on or discussing blockchain technologies and NFTs appear to be limited, with no clear evidence of growth over time.

This study does not explicitly address several additional thematic areas, which are grouped under a broader category labelled 'Other'. This category includes works focusing on diverse artistic practices, digital technologies, interactive systems, installations, games, music, visual studies, cultural heritage, performing arts, as well as studies related to social media or employing social media platforms, among other aspects of the creative domain. Over the examined period, the findings indicate that the Other category exhibits a slight reduction, following the increase in AI-related contributions. Taken together, these results suggest that the growing prominence of AI-related work may be accompanied by a relative compression of topics grouped under the Other category.

The paper is structured as follows. Subsection \ref{sec:related-word} reviews the related work, while Subsection \ref{sec:motivation-objectives-rqs} outlines the motivation and objectives of the study, followed by the research questions (RQs). The Materials and Methods section (\ref{sec:materials-methods}) describes the methodological approach, and is followed by the Results (\ref{sec:results}), Discussion (\ref{sec:discussion}), and Conclusion (\ref{sec:conclusion}) sections.

\subsection{Related Work} 

\label{sec:related-word}
The use of digital technologies in the arts has been widely discussed and has attracted sustained interest from both scholars and artists for many years \cite{tykhoniuk2024analysis, kakunguluintersection}. Beyond these extensive studies, contemporary art and research are largely driven by continuous technological advancements.

Recent studies have systematically examined the use of artificial intelligence in the arts \cite{oksanen2023artificial} and explored its potential impact on artistic practices and art reception \cite{then2023impact, meng2025tracing}. In contrast, systematic approaches comparable to those found in AI-focused studies are largely absent from the literature on extended reality (XR) and its constituent technologies (VR/AR/MR) \cite{wang2023research}, while studies regarding blockchain and NFTs are mostly focused on the art market \cite{lazzaro2022blockchain, abbate2022blockchain} and ownership issues \cite{mochram2022systematic}.

In a relevant systematic review \cite{oksanen2023artificial}, authors investigated empirical studies on the use of AI in fine arts by screening 723 articles and finaly analyzing 44 of them. Through their analysis, they showed that artists use AI mostly for visual arts, followed by music, literature, architecture, and even pottery making, art installation, perfoming arts such as dance and improvised poetry. They also analyzed a significant number of works focusing on human perception of AI in art and identified applications for analysis and classification of artworks. Moreover, their work highlights the steady increase in the number of publications each year. As stated by the authors, although this study covered a rather large range of types of articles, they recognized that their methodological choices might have limited the selection of articles, and, thus, the review might not include all the articles in the field.

In \cite{then2023impact}, authors investigate the impact of artificial intelligence on arts through a systematic review analyzing 30 works of empirical 
research. Among their findings, they identify three (3) main themes: (i) AI’s use in art creation, (ii)
its impact on art appreciation and consumption, and (iii) its ethical
and social implications. They also highlight that according to the analyzed works, more than 70\% (7/10) of the artists
surveyed use AI algorithms in some capacity, from the creation of the initial sketch to the completion of the artwork. As a result a new category of art called ``AI art'' has emerged, so much so that a significant portion of works of art in museums and contemporary art exhibitions are now created, completely or partially, by AI models, and that researchers are using AI to analyze patterns, themes, and techniques in works
of art, enabling a better understanding
of artistic pieces.

In \cite{meng2025tracing}, the authors present a five-year (2021-2025) longitudinal mixed-methods study of 17 Chinese digital painters, examining how their attitudes and practices evolved in response to generative AI. As the authors state, many studies emphasize the opportunities that
AI brings to creative workflows, however, a parallel body of work documents artists’ ambivalence
and resistance. The latter is also a topic of investigation in the considered research \cite{meng2025tracing} with the results showing a progression from resistance and defensiveness, to pragmatic adoption,
and ultimately to reflective reconstruction. The same research also identifies peer influence and personal emotional experience as important factors driving adoption. Regarding its limitations, the authors note the relatively small sample size (17 artists) and the limited geographical and cultural diversity of the participants, all of whom are Chinese. Moreover, its focus on digital visual artists raises questions regarding the reception of AI in other creative fields. With regard to the main finding concerning artists’ shifts in attitude and adoption, the authors highlight that these changes were not disentangled from concurrent technological advances.

\subsection{Motivation and Objectives} \label{sec:motivation-objectives-rqs}

As the analysis of the related work shows (Section \ref{sec:related-word}), the use of emerging technologies is primarily discussed through literature reviews in which the ratio of conference proceedings papers to journal articles is not explicitly defined \cite{oksanen2023artificial}, or where the number of included conference papers is relatively limited \cite{then2023impact}. Other studies \cite{meng2025tracing} adopt mixed-methods approaches that rely mainly on artists as their primary source of information.

Moreover, digital art conferences that are not strongly oriented toward technology may not be indexed in widely used scientific databases such as Scopus or Web of Science. This, in combination with the above factors, can result in such conferences being omitted from literature-based surveys. 

At the same time, conferences play a significant role in the evolution and dissemination of digital arts \cite{fritz2016international}. Compared to journals, conferences enable faster dissemination of research through their proceedings \cite{bjork2019acceptance}, while their typically higher acceptance rates may allow more experimental approaches to reach publication.
Thus, to trace recent shifts in digital arts, this study presents an analysis of discourse in digital art conferences.


Given the aforementioned, the following Research Questions (RQs) are formulated:

\begin{itemize}[ {}]
  \item \textbf{RQ1} (Technological Evolution and Discourse Shift): How do the dominant themes in conference discourse shift over time (2021–2025) in relation to major technological waves, such as immersive technologies (XR, metaverse), Artificial Intelligence, and blockchain/NFTs? \par
  \item \textbf{RQ2} (AI Adoption): When does artificial intelligence (AI) become a central topic in conference contributions, and what types of AI are discussed (e.g., machine learning, deep learning, generative models, and large language models or agents)?
\end{itemize}
\bigskip
\section{Materials and Methods} \label{sec:materials-methods}

This section describes the conference selection process, the filtering applied to conference contributions, and the procedures used for their categorization and analysis.

\subsection{Materials}
Despite the existence of numerous well-known conferences dedicated to digital arts, we avoided selecting conferences based solely on our prior knowledge in order to minimize selection bias. In addition, there is no single focal point that can serve as a common reference, nor is there a comprehensive indexing service—such as those available in other domains (e.g., Scopus or Web of Science)—that facilitates the search and retrieval of digital art conference contributions based on predefined criteria.

Alternatively, the well-known platform WikiCFP is used, as it hosts calls for papers (CFPs) for international conferences, along with announcements for workshops, meetings, seminars, and related events, and provides functionality for identifying relevant international conferences. Unlike the ADA, WikiCFP does not conduct any formal quality evaluation of the conferences it lists. Given the above, the selection of the listed conferences will be performed based on a set of selection criteria appropriate for this study.

\subsection{Methods}
To retrieve the list of conferences dedicated to digital art, the category 'digital art' was selected on WikiCFP \footnote{WikiCFP, page listing 'digital art' conferences: http://www.wikicfp.com/cfp/call?conference=digital\%20art, accessed 01/01/2026}.
To be included in the analysis, a conference had to: (i) appear at least once in the compiled list, and (ii) have taken place during the period 2021–2025. In addition, eligible conferences were required to have been held consistently for the majority of years within the selected time frame (2021–2025).

According to the aforementioned criteria, six (6) conferences are eligible for screening\footnote{Ordered by submission deadline (from most recent to oldest), as listed on WikiCFP.}:
\begin{enumerate}
    \item NICOGRAPH International (NICOINT)
    \item EAI International Conference: Arts, Interactivity \& Game Creation (ArtsIT)
    \item Conference on Electronic Visualisation and the Arts (EVA London)
    \item International Symposium on Electronic Art (ISEA)
    \item Digital Culture \& AudioVisual Challenges: Interdisciplinary Creativity in Arts and Technology (DCAC)
    \item xCoAx: International Conference on Computation, Communication, Aesthetics and X
\end{enumerate}

In addition, one conference (Art Machines 2: International Symposium on Machine Learning and Art, 2021) initially met the two main inclusion criteria; however, it was later identified that its organization had been discontinued and was therefore excluded from the analysis.

Following the identification of eligible conferences, a five-year temporal window (2021–2025) is applied to the selection of submitted contributions, ensuring a focus on the most recent five years. The majority of the eligible conferences are held annually within the selected time frame; the only exception is ISEA 2021, which was postponed to 2022 \cite{isea2022editorial}.

Since the contents of the selected conferences are not uniformly indexed in major scientific databases, the relevant data were manually collected from official conference websites and proceedings repositories (Appendix~\ref{appendix-1}) for the purposes of this study. Conference programmes served as the primary data source and were used to analyse paper titles, session structures, special tracks, keywords, and abstracts when available. Only paper presentations were retained for analysis, while posters, keynotes, panels, and other non-paper contributions were excluded. Overall, a total of 1460 papers were included in the dataset, and their annual distribution per conference is presented in Table~\ref{table:annual-conference-list}.

\begin{table}[h!]
\begin{tabular}{lllllll}
\textbf{Conference / Year} & \textbf{2021} & \textbf{2022} & \textbf{2023} & \textbf{2024} & \textbf{2025} & \textbf{Total (per conference)} \\
\textbf{ISEA}              &  0   &  167    &  21    &  160    & 180   & 528  \\
\textbf{DCAC}              &  74    &  76    &  67    &  43    &  42  &302  \\
\textbf{EVA London}        &  43    &  41    &  39    &  65    & 51   & 239 \\
\textbf{ArtsIT}            &  31  & 46    & 48    &  33    &  56   & 214 \\
\textbf{xCoAx}             &   23    &   15    &    18   &    19   &     19 & 94\\
\textbf{NICOINT}           & 20 &  23    &  12    &  14    & 14 & 83    \\
\textbf{Total (annual)} & 191 & 368 & 205 &334 &362 & 1460
\end{tabular}

\caption{Annual number of paper presentations per conference (2021–2025).}
\label{table:annual-conference-list}
\end{table}

 At the time of analysis, some papers within the dataset may not yet have been formally published at the time of analysis. Consequently, due to limitations in the availability and consistency of abstracts and full-text papers, the analysis was conducted in two stages. In the first stage, the titles of all papers in the dataset were analyzed. In the second stage, a randomly selected subset of papers was examined in greater detail using the available full texts or, when these were not accessible, their corresponding abstracts.

This limitation introduces additional challenges, as text-mining techniques required for clustering and co-occurrence analysis typically depend on relatively large textual datasets \cite{makagonov2004clustering}, while many bibliometric analysis approaches and tools, such as VOSviewer, rely heavily on the availability of well-curated bibliographic data \cite{van2014visualizing}.
Due to the aforementioned constraints, the contributions were manually scanned, identified, and tagged accordingly.

Initially, three main technological categories were defined: (C1) artificial intelligence (AI), (C2) immersive technologies and applications (XR/Metaverse), and (C3) blockchain/NFTs, each accompanied by sets of commonly associated keywords manually identified through extensive screening of conference titles (Table~\ref{tab:categories_keywords}). Moreover, an abstract category called ``Other'' (C4) is created to categorize the contributions that do not fall within the main technological categories.

\begin{table}[h!]
\centering
\small
\begin{tabular}{|p{.09\linewidth}|p{.1\linewidth}|p{.6\linewidth}|p{.1\linewidth}|}
\hline
\textbf{Category ID} & \textbf{Category} & \textbf{Keywords / Tags}  & \textbf{Contri-butions count}\\
\hline
C1 & Artificial Intelligence (AI) &
Artificial Intelligence; AI; Machine Learning (ML); Deep Learning; Computer Vision; Natural Language Processing (NLP);
Language Models; Large Language Models (LLMs); GPT; ChatGPT; Agents; Generative AI; Generative Models; AIGC (AI-Generated Content);
AI-Generated; AI-Generated Art; GAN; Generative Adversarial Networks; StyleGAN2; Diffusion Model(s); Latent Diffusion Model; Stable Diffusion;
Text-to-Image Models; DALL\textperiodcentered E; DALL\textperiodcentered E~3; Midjourney; Prompt Engineering; Prompts;
Reinforcement Learning; Unsupervised Learning; CNN; Convolutional Neural Networks; R-CNN; Swin Transformer; Sentiment Analysis; Support Vector Regression & 206
\\
\hline
C2 & Immersive Technologies and Applications (XR) &
Extended Reality (XR); Immersive Technologies; Immersive; Immersion; Virtual Reality (VR); Augmented Reality (AR); Mixed Reality (MR);
Virtual Environments; Virtual Spaces; Virtual Worlds; Virtuality; Metaverse; Immersive VR; Immersive XR; Immersive Spaces;
Immersive Web; 360-degree Video; Projection Mapping; Holographic; Holographic Installation;
Virtual Museums; Virtual Exhibitions; Virtual Reality Exhibition; Virtual/Augmented Reality; 3D Immersive &  187
\\
\hline

C1 $\cap $ C2 & AI \& XR &
LLM $AND$ VR; Machine Learning $AND$ VR; Generative AI $AND$ Augmented Reality; AI $AND$ Projection Mapping; Artificial Intelligence $AND$ Virtual Reality; Deep Learning $AND$ Augmented Reality; Artificial Intelligence $AND$ Augmented Reality; GAN $AND$ VR & 12
\\
\hline

C3 & Blockchain / NFTs &
Blockchain; NFT; NFTs & 5
\\
\hline

C4 & Other &
Art; Digital; Media; Design;    
Interactive; Human; Data; Game; Technology; Music; Artistic; Visual; Research; Space; Cultural; Practice; Sound; Performance; Learning; Social; Creative; Arts; Heritage; Designing; Body                  & 1072
\\
\hline

\end{tabular}
\caption{Technological categories, associated keywords used for title-based screening, and number of papers identified per category.}
\label{tab:categories_keywords}
\end{table}

Conference papers with titles relevant to at least one of these categories were retained for the subsequent phase of analysis (386 unique titles), while the remaining titles (1072) were excluded. Of the retained titles, 206 fall within the AI category, 187 within immersive technologies (XR/Metaverse), and 5 within blockchain/NFTs, with some titles spanning multiple categories (Table~\ref{tab:categories_keywords}). Moreover, for titles classified under the “Other” category, automatic keyword frequency analysis was selected as the preferred approach due to the volume of the dataset.
Further analysis is discussed in Section \ref{sec:results}.


Given the inherent limitations of the dataset and the limited suitability of automated approaches for accurate topic extraction and clustering, a subset of conference contributions was selected for manual analysis. This subset was determined using simple random sampling in order to reduce selection bias and ensure representativeness.

\bigskip

\section{Results} \label{sec:results}

In this section, the retrieved contributions are analyzed to provide insights into the digital art conference ecosystem, identify indications of shifts in conference discourse, and address the defined research questions. In addition, an analytical overview of a randomly selected set of publications is presented, offering further insight into the AI techniques employed and their application domains.

\subsection{Tracing Shifts in Conference Discourse}

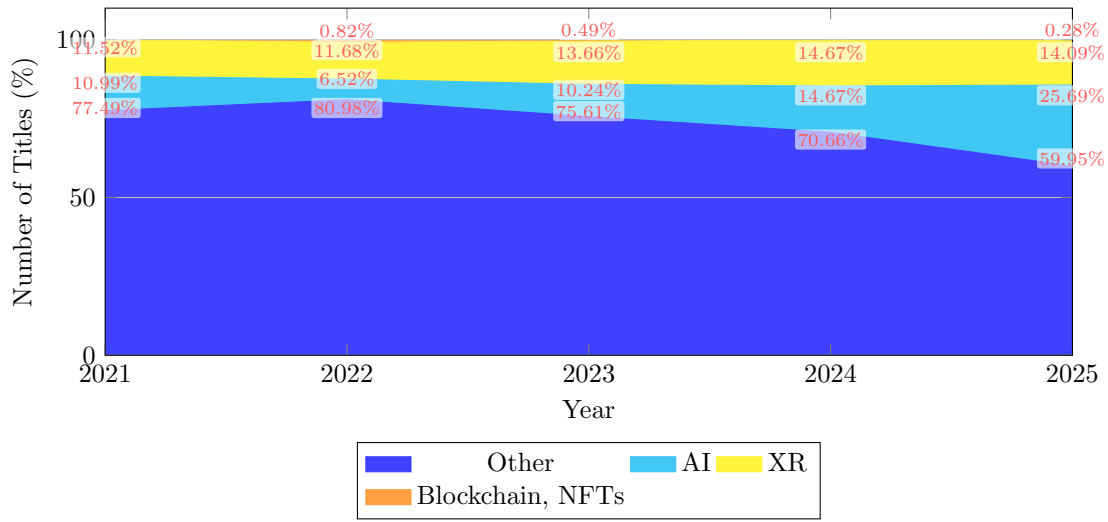
\begin{figure}[t]
\centering
\begin{tikzpicture}
\begin{axis}[
  stack plots=y,
area style,
width=\linewidth,
  height=6.2cm,
enlarge x limits=false,
xlabel={Year},
  ylabel={Number of Titles (\%)},
   ymin=0,
  xtick={2021,2022,2023,2024,2025},
  xticklabel style={/pgf/number format/1000 sep=},
  legend style={at={(0.5,-0.25)}, anchor=north, legend columns=3},
  ymajorgrids=true,
]
\pgfplotsset{
  cycle list={
    blue!75,
    cyan!60,
    yellow!85,
    orange!75,
    red!65,
    blue!45
  }
}

\addplot+[fill] coordinates
    {(2021, 77.49) (2022, 80.98) (2023, 75.61) (2024, 70.66) (2025, 59.95)}
    \closedcycle;
    \addlegendentry{Other}

\addplot+[fill] coordinates
    {(2021,10.99) (2022, 6.52) (2023, 10.24) (2024, 14.67) (2025, 25.69)}
    \closedcycle;
    \addlegendentry{AI}

\addplot+[fill] coordinates
    {(2021,11.52) (2022,11.68) (2023,13.66) (2024,14.67) (2025,14.09)}
    \closedcycle;
    \addlegendentry{XR}

    \addplot+[fill] coordinates
    {(2021,.0) (2022,.82) (2023, .49) (2024,.00) (2025,.28)}
    \closedcycle;
    \addlegendentry{Blockchain, NFTs}


\pgfplotsset{
  every node near coord/.append style={
    font=\scriptsize,
    fill=white, fill opacity=0.6, text opacity=1,
    rounded corners=1pt, inner sep=1pt
  },
  nodes near coords={
    \pgfmathprintnumber[fixed,precision=2]{\pgfplotspointmeta}\%
  }
}

\addplot+[
  only marks, mark=none,
  nodes near coords={\pgfmathprintnumber[fixed,precision=2]{\pgfplotspointmeta}\%},
  point meta=explicit,
  stack plots=false,
  forget plot
] coordinates {
  (2021,75) [77.49]
  (2022,75)  [80.98]
  (2023,74) [75.61]
  (2024,65)  [70.66]
  (2025,59) [59.95]
};

\addplot+[
  only marks, mark=none,
  nodes near coords={\pgfmathprintnumber[fixed,precision=2]{\pgfplotspointmeta}\%},
  point meta=explicit,
  stack plots=false,
  forget plot
] coordinates {
  (2021,82.985) [10.99]
  (2022,84.24)  [6.52]
  (2023,80.73)  [10.24]
  (2024,79) [14.67]
  (2025,79) [25.69]
};

\addplot+[
  only marks, mark=none,
  nodes near coords={\pgfmathprintnumber[fixed,precision=2]{\pgfplotspointmeta}\%},
  point meta=explicit,
  stack plots=false,
  forget plot
] coordinates {
  (2021,94.24)  [11.52]
  (2022,93.34)  [11.68]
  (2023,92.68)  [13.66]
  (2024,92.665) [14.67]
  (2025,92.685) [14.09]
};

\addplot+[
  only marks, mark=none,
  nodes near coords={\pgfmathprintnumber[fixed,precision=2]{\pgfplotspointmeta}\%},
  point meta=explicit,
  stack plots=false,
  forget plot
] coordinates {
  (2022,99.59)  [0.82]
  (2023,99.755) [0.49]
  (2025,99.87)  [0.28]
};

\end{axis}
\end{tikzpicture}
\caption{Stacked area chart showing the distribution of topics per year (2021--2025).}
\label{fig:stacked_topics_year}
\end{figure}

With respect to the evolution of topics within the selected time frame, the majority of conference contributions address themes outside the three focal technological categories (AI, XR, and blockchain/NFTs) as seen in Figure~\ref{fig:stacked_topics_year}. During the early years of the dataset (2021–2023), AI-related contributions remain relatively limited (9.25\% on average), show a slight increase in 2024 (14.67\%), and rise markedly in 2025 (25.69\%). In contrast, XR-related contributions remain comparatively stable across the examined period (minimum: 11.52\% in 2021; maximum: 14.67\% in 2024), exhibiting only minor year-to-year fluctuations. Contributions focusing on blockchain and NFTs are consistently sparse, with no identified contributions in some years (2021 and 2024) and a peak share of 0.82\% in 2022.


With respect to RQ~1, AI-related discourse gains increasing attention over time, as reflected by a rise in the percentage of relevant contributions, while the proportion of contributions categorized as “other” correspondingly decreases. In contrast, XR-related contributions account for a smaller share and remain relatively stable throughout the examined period, whereas blockchain and NFT-related contributions appear only marginally.



With respect to RQ~2, AI does not constitute a central topic in absolute numbers; however, given the high heterogeneity of contributions falling outside the three main technological categories, it can nevertheless be characterized as a central thematic focus.

Further analysis was conducted within the AI category, focusing on the use of specific tools and approaches that are explicitly stated in the titles of the contributions. The relevant keywords (Table~\ref{tab:categories_keywords}) were clustered into seven subcategories: (i) Machine Learning (ML), (ii) Deep Learning (DL), (iii) Large Language Models (LLMs), (iv) Generative Adversarial Networks (GANs), (v) Diffusion Models, (vi) Generative Pre-trained Transformers (GPTs), and (vii) Convolutional Neural Networks (CNNs). 
These subcategories are not mutually exclusive and, in several cases, overlap. For example, CNNs are a class of DL algorithms, while GANs also rely on deep neural networks and therefore are part of the DL family as well. Variants such as Deep Convolutional GANs (DC-GANs) explicitly employ CNN architectures, and GPTs describe transformer-based LLMs that utilize DL techniques to generate human-like text \cite{10521640, goodfellow2014generativeadversarialnetworks}. Nevertheless, these subcategories were retained as separate analytical classes in order to reflect how authors explicitly frame and describe the employed techniques in their paper titles.

\begin{figure}[t]
\centering
\begin{tikzpicture}
\begin{groupplot}[
  group style={
    group size=3 by 3,
    horizontal sep=1.2cm,
    vertical sep=2.0cm
  },
  width=0.3\linewidth,
  height=3.5cm,
  xmin=2021, xmax=2025,
  ymin=0, ymax=3,
  xtick={2021,2022,2023,2024,2025},
  xticklabel style={/pgf/number format/1000 sep=, rotate=45, anchor=east, yshift=-6pt
},
  ymajorgrids=true,
  xlabel={},
  ylabel={},
]

\nextgroupplot[title={Machine Learning}]
\addplot+[mark=*] coordinates {(2021,1) (2022,3) (2023,2) (2024,0) (2025,3)};

\nextgroupplot[title={Deep Learning}]
\addplot+[mark=*] coordinates {(2021,2) (2022,3) (2023,1) (2024,1) (2025,1)};

\nextgroupplot[title={LLMs}]
\addplot+[mark=*] coordinates {(2021,0) (2022,0) (2023,1) (2024,2) (2025,2)};

\nextgroupplot[title={GANs}]
\addplot+[mark=*] coordinates {(2021,2) (2022,1) (2023,0) (2024,0) (2025,1)};

\nextgroupplot[title={Diffusion Models}]
\addplot+[mark=*] coordinates {(2021,0) (2022,0) (2023,1) (2024,1) (2025,1)};

\nextgroupplot[title={GPTs}]
\addplot+[mark=*] coordinates {(2021,0) (2022,0) (2023,0) (2024,0) (2025,3)};

\nextgroupplot[title={CNNs}]
\addplot+[mark=*] coordinates {(2021,1) (2022,0) (2023,1) (2024,0) (2025,0)};

\nextgroupplot[axis lines=none, ticks=none]
\nextgroupplot[axis lines=none, ticks=none]

\end{groupplot}
\end{tikzpicture}

\caption{Temporal distribution of specific AI techniques referenced in the analyzed contributions (2021--2025).}
\label{fig:ai_techniques_over_time}
\end{figure}
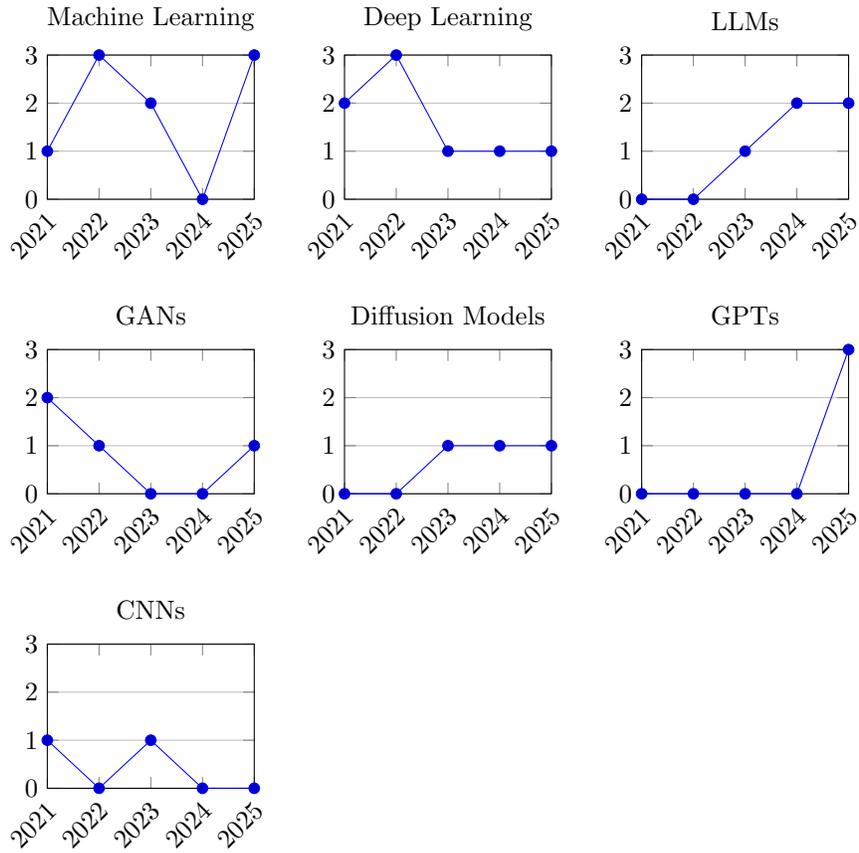

It is observed that the absolute number of contributions explicitly employing these techniques is relatively small and exhibits significant fluctuations over time (Fig.~\ref{fig:ai_techniques_over_time}); nevertheless, their presence helps to identify the main driving forces within the AI category and contributes to partially addressing RQ~2. It is noted, that these contributions represent a small portion of the overall AI category (16\%), as most contributions employ more general terms such as “AI” or “Generative AI.”

\subsection{Distribution of Specific AI Techniques within the AI Category}

Given the above, a sample of contributions employing these more general terms was selected for further analysis in order to identify the use of specific AI techniques. Simple random sampling was employed to select a subset from this group, which consists of 148 titles. The sample size was set to 10\% of the dataset, resulting in fifteen titles (N = 15), which are listed in Table~\ref{tab:selected-ai-titles}. For these contributions, full-text versions were retrieved when available; otherwise, the corresponding abstracts were used for analysis.
Employed techniques and tools were identified; when commercial tools were referenced, their underlying techniques were recorded either as explicitly disclosed in the contribution itself or, when not stated, through available scientific literature, where such information could be reliably identified.

\newpage
\small
\begin{longtable}{ |p{0.05\linewidth}  |p{.22\linewidth}|  p{.22\linewidth}| p{.2\linewidth}| p{.08\linewidth} |p{.05\linewidth}|}
\hline
\textbf{ID - Ref.}  & \textbf{Title}  & \textbf{Techniques / tools}  & \textbf{Topic}  & \textbf{Venue} & \textbf{Year}  \\
\hline

AI.1 - \cite{dcac2021_ai_circe}  & AI CIRCE / ARTIFICIAL INTELLIGENCE AND PAINTING  &  Deep Learning, GAN  & Painting, sentiment analysis, image-to-text  & DCAC & 2021   \\
\hline

AI.2 - \cite{dcac2022_female_spectatorship}  & Female Spectatorship and Artificial Intelligence: The Reversed Gaze in Maria Schrader’s 2021 Film I’m your Man  & Not defined  & Film studies  & DCAC & 2022  \\
\hline

AI.3 - \cite{artsit2022_museum_ai}  & Engaging Museum Visitors with AI-Generated Narration and Gameplay  & NLP, LLMs (GPT-3), Artificial Neural Network (for translation), R-CNN, Object Detector Model (YOLOv5 - CNN \cite{jiang2022review})  & Museums \& museum experience  & ArtsIT & 2022  \\
\hline

AI.4 - \cite{evalondon2022_storytelling_vr}  & Storytelling and VR: Inducingemotions through AI characters  & Behavior Trees, Utility AI (games)  & Video Game development  & EVA London & 2022  \\
\hline

AI.5 - \cite{xcoax2022_hyperobject_portal}  & AI Art as Hyperobject-Like Portal to Global Warming  & Not explicitly defined  & Ecological awareness / global warming & xCoAx & 2022  \\
\hline

AI.6 - \cite{xcoax2022_lures_engagement}  & Lures of Engagement: An Outlook on Tactical AI Art  & Machine Learning, Deep Learning, NLP,  GAN , CNN   & AI Art, Chatbots, Computer Vision, text-to-image, image-to-text,  & xCoAx & 2022  \\
\hline

AI.7 - \cite{xcoax2023_haunted_ai}  & ‘Haunted’ AI  & Diffusion Models (Dall-E) & Text-to-image, visual culture  &  xCoAx & 2023  \\
\hline

AI.8 - \cite{dcac2023_script_pitching}  & From the End to the Beginning of a script : Pitching essentials in AI era  & Not defined  & Cinema, movie scripts  & DCAC & 2023  \\
\hline

AI.9 - \cite{isea2024_animation_ai}  & Animation in the Age of AI: Creative Dialog With Algorithms  &  Kaiber.ai4, Runway Gen25 (Latent Diffusion Model \cite{rombach2022high}), Midjourney6 (Transformer \cite{derevyanko2023comparative}),
Stable Diffusion XL7 (Diffusion Model), Automatic11118 (Stable Diffusion / Diffusion model \cite{brade2023promptify})  & AI-driven Image Generation, computer animation, animated short film  & ISEA & 2024  \\
\hline

AI.10 - \cite{evalondon2024_immersive_ai_language}  & Immersive AI-Driven Language Learning: animating languages through gamified encounters  & LLMs (ChatGPT) & Gamified Language-learning  & EVA London & 2024  \\
\hline

AI.11 - \cite{evalondon2024_technoetic_magick}  & Technoetic Magick: Explorations of
the Uncanny Double as a Noetic and
Magickal System through the
Complementary Lenses of AI Image
Generation and AR  & Playform (AI Art Generator), no other techniques are defined  & Image Generation  & EVA London & 2024 \\
\hline

AI.12 - \cite{isea2025_composable_life}  & Composable Life: Speculation for Decentralized Al Life  & Not defined  & Science fiction  & ISEA & 2025  \\
\hline

AI.13 - \cite{dcac2025_position_paper}  & Position Paper: Creating and Managing Funded Cultural Projects through Artificial Intelligence and Big Data  &  Not defined  & Grant proposals, project management   & DCAC & 2025  \\
\hline

AI.14 - \cite{isea2025_enhancing_immersive_art}  & Enhancing lmmersive Art Appreciation through Al-Generated
Personalized Stories  & Not defined  & Art appreciation  & ISEA & 2025  \\
\hline

AI.15 - \cite{isea2025_bada_shanren}  & Bada Shanren's Dadi Caotang: Exploring Virtual Reconstruction of a Lost Chinese Painting with Generative Artificial Intelligence  & Stable Diffusion  & Text-to-image, image reconstruction  & ISEA & 2025  \\
\hline
\caption{Identification of Employed AI Techniques, Tools, and Application Topics in a Random Sample of Contributions Using General Terms such as “AI” or “Generative AI”.} \label{tab:selected-ai-titles} 
\end{longtable}

\begin{figure}[h]
\centering
\begin{tikzpicture}
\begin{axis}[
  ybar,
  width=0.7\linewidth,
  height=6cm,
  ylabel={Number of contributions},
  symbolic x coords={
    ML,
    DL,
    LLMs,
    GANs,
    Diffusion models,
    GPTs,
    CNNs,
    Transformers,
    Behavior trees
  },
  xtick=data,
  xticklabel style={rotate=45, anchor=east},
  ymin=0,
  ymajorgrids=true,
  bar width=18pt,
  nodes near coords,
  nodes near coords align={vertical},
]
\addplot coordinates {
  (ML, 1)
  (DL, 2)
  (LLMs, 2)
  (GANs, 2)
  (Diffusion models, 2)
  (GPTs, 2)
  (CNNs, 2)
  (Transformers, 1)
  (Behavior trees, 1)
};
\end{axis}
\end{tikzpicture}
\caption{Distribution of AI techniques identified in the analyzed sample of contributions employing general AI terminology.}
\label{fig:ai_techniques_bar}
\end{figure}
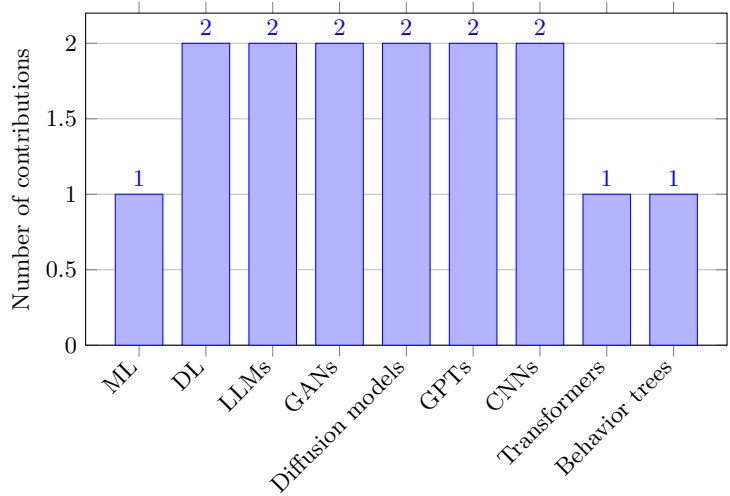

The analysis of the sample drawn from contributions employing more general AI terminology confirms the previous findings regarding the most frequently employed techniques and introduces two additional techniques not previously identified: (i) Transformers, which are closely related to GPTs, and (ii) Behavior Trees \cite{shoulson2011parameterizing, gugliermo2024evaluating} (Fig.~\ref{fig:ai_techniques_bar}).

In a substantial number of contributions, the employed techniques are not explicitly specified. This can be attributed either to a focus on theoretical or conceptual aspects, or to the fact that the work under discussion has not yet matured sufficiently to provide detailed technical information.

\bigskip
\section{Discussion} \label{sec:discussion}

The findings indicate an increase in contributions that use artificial intelligence for creative purposes, which is also confirmed by other works \cite{oksanen2023artificial}. The observed increase can be attributed to the fact that the AI market witnessed some major events during this period. Specifically, ChatGPT  (GPT 3.5 - LLM)\footnote{OpenAI, Introducing ChatGPT: https://openai.com/index/chatgpt/ (accessed 20/01/2026)}, DALL·E 2 (Diffusion model)\footnote{OpenAI, DALL·E 2 research preview update, https://openai.com/index/dall-e-2-update/ (accessed 20/01/2026)}, MidJourney\cite{tsidylo2023artificial}, and Stable Diffusion\cite{tsidylo2023artificial}, were announced and became widely available on 2022. Indeed, before 2023, only a small number of the analyzed contributions referenced the use of LLMs, GPTs, and  Diffusion Models, but 2023 indicates an increase on these references.

Based on the widespread use of AI in the examined works and the increasing number of related contributions, it appears that recent developments in AI have generated a growing sense of enthusiasm within the digital art scene. Despite the overall positive attitude toward AI technologies, the examined dataset also reveals voices of concern regarding their high ecological costs \cite{huang2025isea_ecologycomplexities}. At the same time, other works \cite{xcoax2022_hyperobject_portal} employ AI-based artistic practices to raise awareness of contemporary ecological challenges, such as global warming.

Immersive technologies—including the family of extended reality (XR) technologies (VR/AR/MR) and the metaverse—constitute the second most prominent topic, following AI-related work, and have remained a relatively stable area of interest throughout the period under study. Despite recent technological progress and market adoption, concerns have been raised that virtual reality (VR) may be entering a so-called “winter” phase, a term commonly used to describe periods of slowed growth \footnote{Meester, Raymond. “Virtual Winters.” Medium, 9 Nov. 2021, https://raymondmeester.medium.com/virtual-winters-1a5690e47c59 (accessed 20/01/2026)}$^,$ \footnote{Evans, Benedict. “The VR Winter Continues.” ben-evans.com, 8 July 2024.
https://www.ben-evans.com/benedictevans/2024/7/8/the-vr-winter-continues (accessed 20/01/2026)}.
~This context may help explain the apparent plateau observed in conference contributions related to VR.
A similar pattern of stabilization or limited expansion can be observed in contributions related to blockchain technologies and NFTs, which remain comparatively scarce and show no clear evidence of growth over time. Overall, it appears that the increasing prominence of AI-related topics may be accompanied by a compression of the Other category, which exhibits a slight decrease over time.


 



\subsection{Limitations}

WikiCFP does not perform any formal quality control on the conferences it indexes. In other research domains, curated catalogues and ranking systems for conferences do exist, such as discipline-specific conference lists and evaluation portals (e.g., CORE Conference Rankings\footnote{CORE Conference Portal. Australian and New Zealand Computer Science Research Rankings, Computing Research and Education Association of Australasia, https://www.core.edu.au/conference-portal (accessed 20/01/2026)}). However, no equivalent, widely accepted classification or ranking framework is currently available for conferences in the field of digital art.

The conferences selected for this study can therefore be considered broadly representative of the current landscape of digital art research and practice. Nevertheless, it is possible that other significant conferences were not identified or included in the dataset.

Furthermore, the manual retrieval and processing of conference data is a time-consuming task, which constrains the achievable sample size. While the analysis of paper titles provides an initial overview of thematic trends, it is not sufficient for deeper qualitative or conceptual analysis. Issues related to sampling—such as saturation, representativeness, and sample size—thus remain important considerations and constitute limitations of the present study.







\bigskip

\section{Conclusion} \label{sec:conclusion}
This study was motivated by the hypothesis that recent advances in AI techniques—together with substantial market investment and subsequent adoption—would be reflected in the discourse of digital art conferences. Moreover, systematic reviews in this area remain scarce and, in many cases, do not adequately capture the dynamics of digital art conferences, which often serve as venues for presenting novel and experimental work within shorter timeframes than journals, given the latter’s higher rejection rates and longer publication cycles.

In this context, the study succeeds in identifying a set of representative conferences that appear to be well established within their respective communities and have been held consistently over a sufficiently long period to be considered long-lived and mature. Following the identification of the examined conferences, the research team retrieved and analyzed the contributions presented in paper form. Initially, all paper titles were screened for keywords indicating the main topics of each work, enabling their categorization. Subsequently, a random sample of contributions classified under the AI topic was examined in greater detail in order to identify the techniques employed as well as the application domains.

With respect to the defined Research Questions (\ref{sec:motivation-objectives-rqs}): \begin{itemize}
    \item RQ-1: AI-related discourse gains increasing attention over time, as reflected by a rise in the percentage of relevant contributions, while the proportion of contributions categorized as “other” correspondingly decreases. In contrast, XR-related contributions account for a smaller share and remain relatively stable throughout the examined period, whereas blockchain and NFT-related contributions appear only marginally.
    \item RQ-2: With respect to RQ~2, AI does not constitute a central topic in absolute numbers; however, given the high heterogeneity of contributions falling outside the three main technological categories, it can nevertheless be characterized as a central thematic focus.
\end{itemize} 

Regarding the other evolving technologies that were also targeted, namely immersive technologies, and blockchain technlogies and NFTs, the findings showed that the former constitutes an important topic of the conference discourse with a significant portion of the contributions that on the one hand remains stable, but, on the other hand, the latter observation can also be characterized as a plateau and signs of the so-called VR winter. In contrast, contributions focusing on blockchain technologies and NFTs are marginal and do not exhibit any sign of growth during the examined period.
In addition, this work provides insights into the digital art conference ecosystem and highlights the need for a more clearly defined and systematically compiled list of digital art conferences.

\section{Disclosure (Use of Generative AI Tools)}
A large language model (ChatGPT, version 5.2, Plus plan, OpenAI) was used to assist with drafting and refining the manuscript text, as well as to support limited data preprocessing tasks. The formulation of the research questions, the overall study design, and the conceptual framing of the work were entirely developed by the authors. The model was not used for data analysis or decision-making; all analytical judgments and interpretations remain the responsibility of the authors. All AI-assisted content was reviewed and edited by the authors, who take full responsibility for the content of the publication.

\bigskip
\section{Acknowledgments}
  This work did not receive any funding.  
\bigskip

\appendixpage
\renewcommand{\thesection}{\Alph{section}}
\setcounter{section}{0}
\section{Conference websites, programs \& proceedings} \label{appendix-1}
\begin{enumerate}

    \item ISEA
    \begin{enumerate}
        \item  2025: https://www.isea-symposium-archives.org/publications/, 	https://www.isea-symposium-archives.org/wp-content/uploads/2025/10/ISEA2025-Proceedings.pdf

        \item 2024: https://www.isea-symposium-archives.org/publications/,	https://www.isea-symposium-archives.org/wp-content/uploads/2025/10/ISEA2024\_Proceedings.pdf

        \item 2023: https://www.isea-symposium-archives.org/publications/,	https://www.isea-symposium-archives.org/wp-content/uploads/2025/09/3rd-Summit-Proceedings\_Final\_Papers\_Version2.0\_01SEPT2023.pdf
    
        \item 2022: https://www.isea-symposium-archives.org/wp-content/uploads/2023/02/ISEA2022-Proceedings\_.pdf
    
        \item 2021: https://www.isea-symposium-archives.org/pres-overview/isea2021-presentation-overview/	(No presentations / papers)
    \end{enumerate}

    \item DCAC 
    \begin{enumerate}
        \item 2025: https://avarts.ionio.gr/dcac/2025/en/schedule/	

        \item 2024:	https://avarts.ionio.gr/dcac/2024/en/schedule/	
        \item 2023:	https://avarts.ionio.gr/dcac/2023/en/schedule/	

        \item 2022:	https://avarts.ionio.gr/dcac/2022/en/schedule/

        \item 2021	https://avarts.ionio.gr/dcac/2021/en/schedule/, https://avarts.ionio.gr/dcac/2021/en/proceedings/
    \end{enumerate}

     \item EVA London 
     
     \begin{enumerate}
         \item 2025:	http://www.eva-london.org/eva-london-2025/, 	http://www.eva-london.org/wp-content/uploads/2025/08/EVA\_London\_2025\_programme\_2025-07-11.pdf

        \item 2024:	http://www.eva-london.org/eva-london-2024/, http://www.eva-london.org/wp-content/uploads/2024/07/EVA\_London\_2024\_programme\_2024-07-12-compressed.pdf
    
        \item 2023:	http://www.eva-london.org/eva-london-2023/programme/, http://www.eva-london.org/wp-content/uploads/2023/07/EVA\_London\_2023\_programme.pdf
    
        \item 2022:	http://www.eva-london.org/eva-london-2022/programme/, http://www.eva-london.org/wp-content/uploads/2022/07/EVA-London-2022-programme.pdf
    
        \item 2021:	http://www.eva-london.org/eva-london-2021/programme/, http://www.eva-london.org/wp-content/uploads/2021/04/EVA-London-2021-presentations.pdf

     \end{enumerate}
     
    \item ArtsIT 
    
    \begin{enumerate}
        \item  2025:	https://artsit.eai-conferences.org/2025/accepted-papers/	

        \item 2024:	https://artsit.eai-conferences.org/2024/accepted-papers/	
    
        \item 2023:	https://artsit.eai-conferences.org/2023/accepted-papers/
    
        \item 2022:	https://artsit.eai-conferences.org/2022/,	https://link.springer.com/book/10.1007/978-3-031-28993-4
        
        \item 2021:	https://artsit.eai-conferences.org/2021/, 	https://artsit.eai-conferences.org/2021/accepted-papers/
    \end{enumerate}

    \item xCoAx	

    \begin{enumerate}
        \item  2025:	https://2025.xcoax.org/,	https://2025.xcoax.org/\#program

        \item 2024:	https://2024.xcoax.org/\#papers

        \item 2023:	https://2023.xcoax.org/\#papers	

     \item2022	https://2022.xcoax.org/html/program.html

        \item 2021:	https://2021.xcoax.org/papers/	
    \end{enumerate}

    \item NICOINT 
    
    \begin{enumerate}
        \item 2025: https://www.art-science.org/nicograph/nicoint2025/\#nicoint2025-program	
    
    \item 2024:	https://www.art-science.org/nicograph/nicoint2024/\#nicoint2024-program

    \item 2023:	https://www.art-science.org/nicograph/nicoint2023/\#nicoint2023-program	

    \item 2022: https://www.art-science.org/nicograph/nicoint2022/\#nicoint2022-program	

    \item 2021	https://www.art-science.org/nicograph/nicoint2021/\#nicoint2021-program	
    \end{enumerate}

\end{enumerate}

\end{document}